\documentclass[9pt,twocolumn,twoside]{opticajnl}
\journal{opticajournal} 

\setboolean{shortarticle}{false}

\usepackage{lineno}
\usepackage{siunitx}
\usepackage{soul}
\usepackage{amsthm}

\title{High-Speed Widefield Hyperspectral Imaging}

\author[1,2,*]{Joana M. Teixeira}
\author[1,2]{Tomás Lopes}
\author[1]{Tiago D. Ferreira}
\author[1]{Catarina S. Monteiro}
\author[1,2]{Pedro A.S. Jorge}
\author[1,2]{Nuno A. Silva}

\affil[1]{Centre for Applied Photonics, INESC TEC, Rua do Campo Alegre 687, Porto, 4169-007, Porto, Portugal.}
\affil[2]{Departamento de Física e Astronomia, Faculdade de Ciências, Universidade do Porto, Rua do Campo Alegre s/n, Porto, 4169-007, Porto, Portugal.}

\affil[*]{joana.m.teixeira@inesctec.pt}

\begin{abstract}
Fourier-transform hyperspectral imaging is an emerging technology that combines widefield acquisition with flexible spectral resolution. Yet, it typically requires the acquisition of dense sequences of full-field interferograms during optical-path scanning, which ultimately reduces the final throughput. To circumvent this limitation, this manuscript introduces an event-based Fourier-transform hyperspectral microscopy technique that exploits the coupling of a neuromorphic event camera with a common-path birefringent interferometer operated under continuous scanning. Specifically, two complementary acquisition strategies are investigated, one with steady broadband illumination where the event stream corresponds to a derivative-like interferogram, and another with a temporal modulation of a broadband LED, which produces an intensity-like event response. Both approaches are able to retrieve the principal spectral profiles of photographic filters, showing a good qualitative match with both frame-based and spectrometer references. Compared against the typical 15-minute acquisition time for step-scan reference configuration, both methods represent a speedup larger than 3 orders of magnitude, entering real-time acquisition speeds: the derivative-based method acquires the useful interferometric range in approximately 200 ms, corresponding to an effective rate of approximately 5 hyperspectral cubes per second, whereas the modulation-assisted method achieves an experimentally measured spectral resolution of 17.9 nm at an effective acquisition rate of 2.5 datacubes per second. Finally, reflection-mode imaging and motion-compensated reconstruction of moving particles further demonstrate spatially resolved spectral recovery beyond static transmission measurements, establishing the solution as a promising technology for continuous, wide-field hyperspectral imaging.
\end{abstract}

\setboolean{displaycopyright}{false} 

\begin{document}

\maketitle

\section{Introduction}
Hyperspectral Imaging (HSI) is an optical sensing technique that combines spectroscopy and photography into a single system, capturing spatially resolved images across numerous contiguous spectral bands. In fact, these systems produce pixel-level spectra that are assembled into three-dimensional datacubes, containing information regarding the reflectance/radiance of the sample and enabling the characterization of materials, tissues, and even large surfaces in environmental monitoring scenarios \cite{prabhavathy2019unsupervised, maktabi2025detection, giannoni2026hyperspectral, pahlawan2025non, zhang2025monitoring}.  HSI focuses on the optical window of the electromagnetic spectrum, typically covering wavelengths from 380 to 2500 nm in most standard configurations, encompassing both the visible region as well as the near-infrared and short-wave infrared regions \cite{hong2026hyperspectral} with spectral resolutions or sampling intervals on the order of 5–10 nm \cite{Sinclair_06, chabrillat2024enmap}. Alternatively, in simpler, lower-cost systems, the full spectrum is divided in fewer spectral bands, typically around 20 or so. The former are designated hyperspectral systems, while the latter are more often classified as multispectral. 

The resulting three-dimensional data cubes provide detailed spectral information at each spatial location, supporting label-free and non-destructive analysis of the observed sample. However, mapping a three-dimensional $(x,y,\lambda)$ hyperspectral datacube onto a two-dimensional detector array introduces fundamental trade-offs in system design and data acquisition. Conventional HSI architectures rely primarily on spatial or spectral scanning. Point-scanning, also known as whiskbroom, systems sequentially capture individual pixels using mechanically actuated mirrors. However, their mechanical complexity results in slow acquisition speeds. In fact, depending on the scanned area and spatial sampling, acquisitions may require from several minutes to several hours \cite{holman2020autonomous}. Similarly, Line-scanning (pushbroom) systems capture spatial lines simultaneously and build a two-dimensional image as the platform moves forward. This substantially reduces acquisition time, although complete data cubes may still require several seconds, with examples showing the acquisition of 100 spectral bands in  5-10 seconds \cite{von2025image}. Alternatively, scanners can use electronically tunable filters to acquire spectral bands sequentially. These extended acquisition times require highly stable measurement conditions, severely limiting their applicability to dynamic or rapidly changing scenes. Although the exact performance depends on the instrument configuration, all three approaches acquire part of the datacube sequentially, making them sensitive to motion and illumination changes and limiting their applicability to dynamic scenes. Snapshot HSI circumvents mechanical scanning by acquiring spatial and spectral information simultaneously within a single exposure, enabling substantially faster imaging of dynamic scenes. For example, Gao et al. \cite{gao2011snapshot} reported an image-mapping spectrometer that acquired 48 spectral bands, with a spectral bandwidth of approximately 4\,nm, at 5.2 frames per second. However, this temporal advantage is often obtained at the expense of spatial sampling, which in this system was limited to 350$\times$350 pixels because the detector area must accommodate both spatial and spectral information.

Fourier Transform Hyperspectral Imaging (FT-HSI) offers an elegant alternative to spatial multiplexing techniques. By pairing a widefield sensor with a two-beam interferometer, it encodes spectral information in the temporal domain as an interferogram, preserving the native spatial resolution of the detector \cite{puckrin2009airborne}. Extracting accurate spectra requires scanning the optical path difference at precise micro-steps, which, when utilizing conventional complementary metal-oxide-semiconductor (CMOS) or charge-coupled device (CCD) sensors, demands capturing thousands of full-field, dense frames per datacube. Because HSI captures hundreds of contiguous spectral bands, the resulting data volumes are several orders of magnitude larger than standard imaging, creating severe memory, bandwidth, and storage bottlenecks. These data transmission and readout limits strongly constrain the capacity of conventional FT-HSI to achieve simultaneous high spatial resolution and high acquisition speed.

Neuromorphic event-based vision sensors (EVS) provide a hardware-level solution to these bandwidth limitations. Inspired by biological retinas, event cameras contain independent pixels that operate asynchronously, generating data only when local changes in logarithmic intensity exceed a predefined threshold \cite{lichtsteiner2008}. This sparse, event-driven readout reduces redundant temporal sampling, yielding microsecond precision, extremely high dynamic ranges, and heavily compressed data streams \cite{ gallego2020}. Motivated by these advantages, recent studies have begun integrating event cameras into spectral and multidimensional imaging systems \cite{zhang2024event,guo2024eventlfm,teixeira2026asynchronous,geckeler2026event}. Yu et al. \cite{yu2025active} developed an active event-based hyperspectral imager in which synchronized, spectrally swept illumination encodes wavelength-dependent variations as temporal events, enabling capture at up to 30 frames per second while reducing the reported data bandwidth. More recently, Chen et al. \cite{chen2026self} introduced OpenHI, a transmission-mode hyperspectral microscope that mechanically sweeps dispersed illumination across the specimen and reconstructs spectral derivatives of the sample’s log-transmittance from the resulting event stream. Baird et al. \cite{baird2025dynamic} instead employed a hybrid event–CMOS architecture, using event sensing for high-speed particle tracking and a diffractive CMOS channel for spectral discrimination. Together, these studies demonstrate the potential of event-based sensing for high-speed spectral and multidimensional imaging. Nevertheless, current event-based hyperspectral systems generally rely on actively or mechanically swept illumination to generate wavelength-dependent temporal contrast, whereas hybrid approaches may require an additional frame-based detector, limiting fully passive, single-sensor wide-field operation.

Beyond these peer-reviewed implementations, a related architecture combining an event camera with a Fourier-transform interferometer, including a TWINS implementation, has previously been described in a patent \cite{ritter2024hyperspectral}. The proposed approach aggregates events over temporospatial neighborhoods and applies demodulation followed by a frequency-domain transform, such as a non-uniform discrete cosine transform, to estimate the optical spectrum. The disclosure does not, however, report an experimental implementation or measured hyperspectral reconstructions. In particular, it does not experimentally characterize the derivative-like event response produced during continuous interferometer scanning or its compensation during spectral reconstruction, nor does it consider the illumination-modulation strategy investigated here for obtaining an intensity-like event response. To the best of our knowledge, no peer-reviewed experimental demonstration and characterization of this interferometric event-based architecture has previously been reported.

In this work, we experimentally demonstrate and characterize a wide-field interferometric framework that directly couples an event-based camera with a common-path birefringent interferometer. In contrast to event-based hyperspectral approaches that encode wavelength by spectrally sweeping the illumination across the sample, the complete broadband image is preserved throughout the acquisition, while spectral information is encoded through a continuously varied interferometric delay. The resulting temporal evolution of the interferogram is recorded asynchronously by the event camera, such that only local changes in the interferometric signal are transmitted. This architecture decouples spatial sampling from spectral encoding: the detector area remains entirely available for imaging, while the spectral dimension is carried by the temporal interferometric response rather than being distributed spatially across the sensor or sequentially selected at the sample. The event-based readout therefore complements Fourier-transform hyperspectral imaging particularly well, replacing the dense sequence of full-field interferograms required by conventional frame cameras with a sparse stream of interferometric changes. In this way, the proposed approach combines native wide-field spatial sampling with continuous interferometric spectral encoding and substantially reduced redundant data acquisition.

\section{Methods and Results}

\subsection{Optical Setup}

\begin{figure*}[h!]
    \centering
    \includegraphics[width=0.9\linewidth]{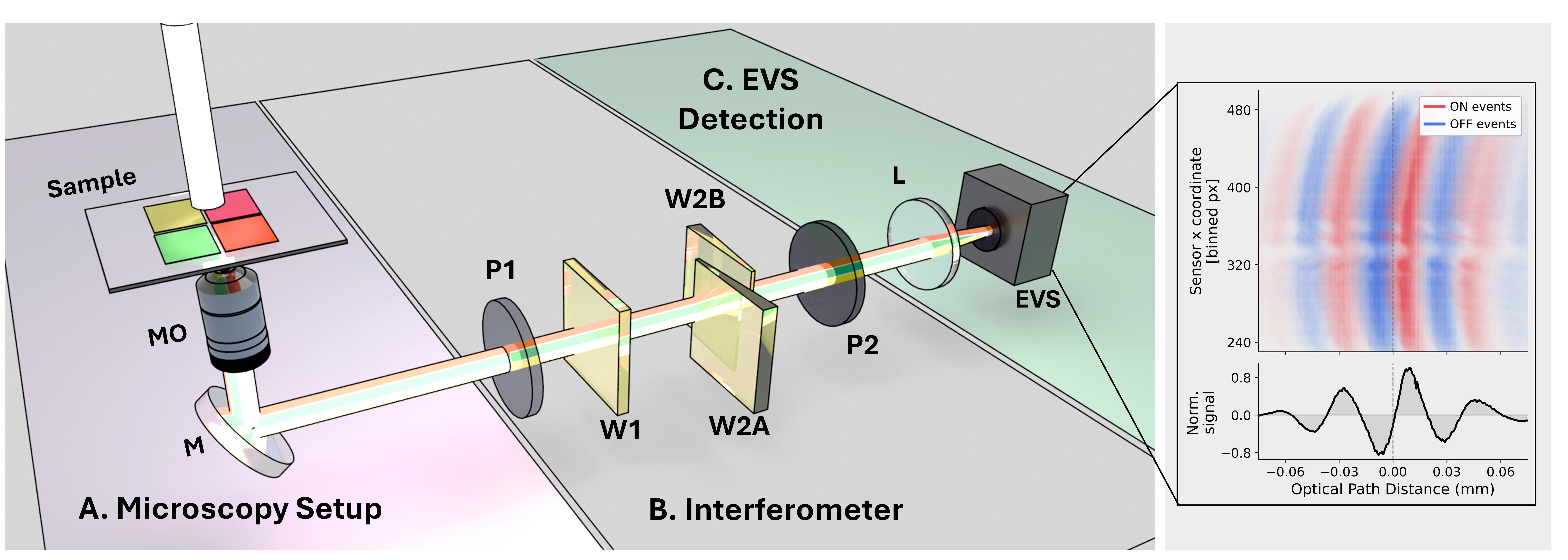}
    \caption{Experimental optical layout used for event-based interferometric spectral imaging. The sample consists of colored photographic filters mounted on a microscope slide and imaged through a 4× microscope objective (MO). The transmitted light is redirected by a mirror (M) into the GEMINI interferometer from NIREOS, composed of a polarizer (P1), birefringent wedges (W1, W2A, and W2B), and an analyzer (P2). The resulting interferogram is relayed by a 50 mm focal-length lens (L) onto the event-based vision sensor (EVS), where fringe motion is recorded asynchronously as positive and negative events. The right panels show the corresponding event cloud as a function of optical path distance and the integrated event-domain interferogram extracted from it.}
    
    \label{fig:setup}
\end{figure*}

The optical setup begins with an inverted microscope configuration (see Figure \ref{fig:setup}A), where the sample is positioned on a glass slide and illuminated in transmission mode using a white LED (see section S1 of the Supplementary Material for the illumination spectrum). The transmitted light is collected by a 4x microscope objective, which was selected to provide a large depth of focus. The resulting collimated beam is then reflected directly into a commercial common-path interferometer (GEMINI, Nireos) for spectral encoding.

The sample used in this work consisted of a combination of LEE photographic filters with well-characterized transmission spectra.\cite{leeFiltersColourEffect}. To fully test the imaging capabilities of our system, we selected four filters with complex spectral profiles: 793 - Vanity Fair, 765 - LEE Yellow, 652 - Urban Sodium, and 728 - Steel Green. All reference transmission spectra were recorded with a broadband spectrometer (AvaSpec-ULS2048CL-EVO, Avantes), using the same illumination source, by integrating over a small sample area.


The GEMINI is a common-path birefringent interferometer based on TWINS (Translating-Wedge-Based Identical Pulses eNcoding System) technology \cite{perri2021visible}. An input polarizer prepares a well-defined polarization state, while the birefringent elements generate two collinear, orthogonally polarized replicas of the incoming field. Translating one of the birefringent wedges varies the relative optical delay, $\tau$, between the two replicas while preserving their common propagation path, providing high intrinsic phase stability. Finally, an analyzer oriented at $45^\circ$ projects the two polarization components onto a common axis, allowing them to interfere before detection.

Exiting the interferometer, the spatially resolved image of the sample is naturally superimposed onto the resulting interference pattern. To capture this signal, a 50\,mm focal length lens images the combined output directly onto the sensor of a 1280x720 pixel event-based camera (EVK4, Prophesee) (Figure \ref{fig:setup}C.). For comparison, a monochrome frame-based camera (CS165MU/M, Thorlabs) was also used, following the same configuration.

\subsection{Frame-Based Reconstruction}

A conventional frame-based acquisition was first performed to provide a reference for the event-based measurements. This was implemented using the standard step-scan operating mode of the GEMINI interferometer, in which the motorized birefringent wedge is sequentially displaced to a set of predefined positions and a complete intensity image is recorded after the wedge reaches each position. The optical configuration and sample illumination is the one shown in Figure \ref{fig:setup}. However, for this reference measurement the event-based sensor was replaced with a monochrome frame camera. The camera was operated using the full sensor area, for which its maximum frame rate is 34.8 fps. Slightly higher frame rates can be obtained by reducing the sensor region of interest or using pixel binning. Nevertheless, we retained full-sensor acquisition in this work to preserve the complete field of view and native spatial sampling, providing an appropriate reference against which to compare the event-based acquisition.

To acquire the hyperspectral datacube, the GEMINI wedge position was scanned from $-0.4$ to $+0.4$~mm over 1000 uniformly distributed positions, corresponding to a step size of approximately $0.80~\mu$m.  At each position, the wedge was firstly moved to the target position, followed by a 1\,ms settling time. A full-field image was then recorded with an exposure time of 0.5\,ms. The scan range was selected to obtain sufficient spectral resolution, since the maximum interferometric excursion determines the achievable spectral resolution in Fourier-transform spectroscopy \cite{perri2021visible}. The dense position sampling also ensures that the interferometric oscillations over the visible spectral range are adequately sampled. Despite the frame rate of the camera, the repeated wedge translation, stabilization, camera readout, and data transfer increase the measurement time, with the full acquisition requiring approximately 15 minutes.

The acquired image sequence can be represented as
$I(x_k,y,z)$, where $x_k$ is the measured GEMINI position and
$(y,z)$ denotes the spatial coordinates on the camera. For each
spatial pixel, the intensity values recorded throughout the scan form
an interferogram. The measured signal can be written as

\begin{equation}
I(x) = I_0 + \int S(\nu)\cos(2\pi\nu x)\,d\nu,
\label{eq:interferogram}
\end{equation}

where $I_0$ is the non-interferometric background and $S(\nu)$ is
the spectral distribution expressed as a function of the calibrated
interferometric spatial frequency $\nu$. Before reconstruction, each frame was associated with its corresponding wedge position and ordered along the position axis. A slowly varying background was
then estimated and subtracted from each pixel interferogram, after
which a Gaussian apodization window was applied. A phase-corrected Fourier transform was subsequently calculated along the GEMINI wedge position dimension. The spectrum at each spatial coordinate was therefore obtained as

\begin{equation}
S(\nu) \propto
\left|
\mathcal{F}_{x}
\left\{
I(x)-I_{\mathrm{bg}}(x)
\right\}
\right|,
\end{equation}

where $I_{\mathrm{bg}}(x)$ denotes the estimated background. The
resulting interferometric spatial-frequency axis was converted to
wavelength, producing the hyperspectral datacube
$S(y,z,\lambda)$. The wavelength calibration of the GEMINI interferometer was performed by the manufacturer and provided as a calibration file.

\begin{figure*}[h!]
    \centering
    \includegraphics[width=0.9\linewidth]{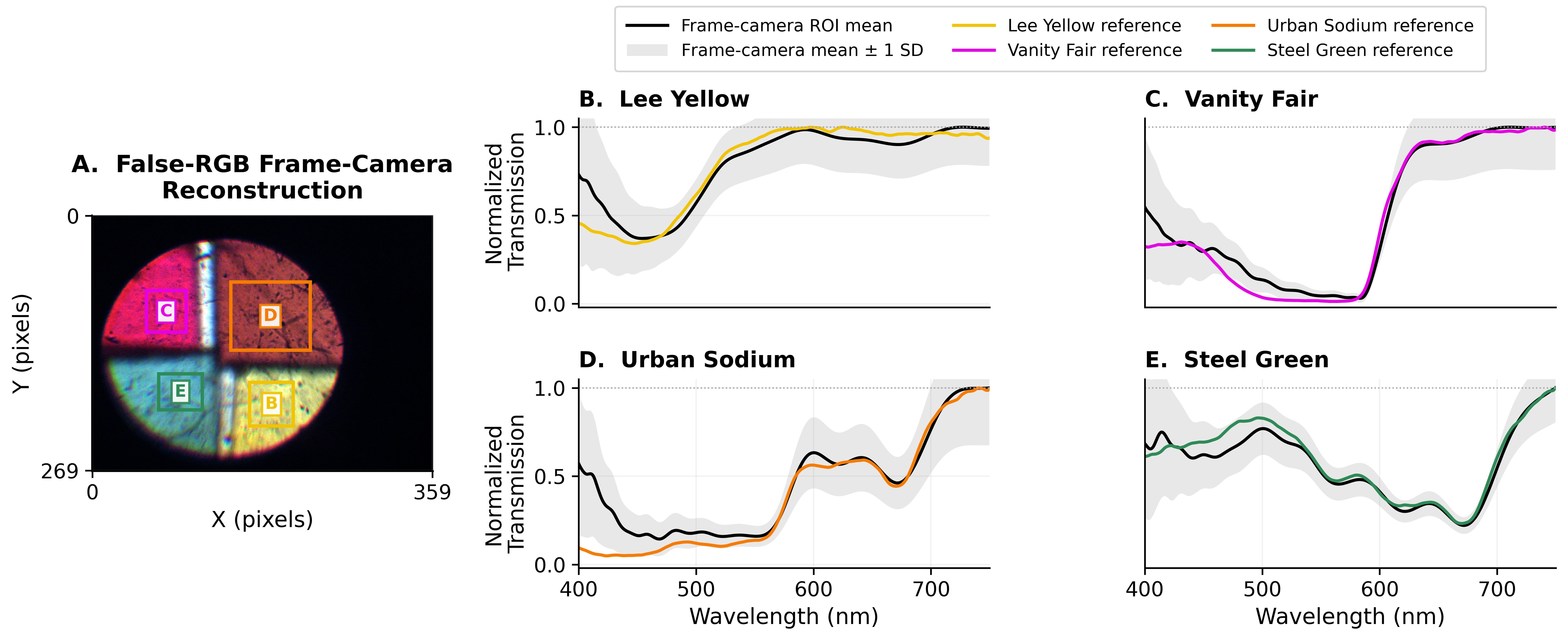}
    \caption{Frame-based reference hyperspectral reconstruction.
    (A) False-color representation of the sample generated by averaging the reconstructed datacube over the blue (440-470~nm), green (510-560~nm), and red (620-700~nm) spectral intervals. The labeled markers indicate the regions from which the spectra were extracted. (B-E) Reconstructed transmission spectra for LEE
    Yellow, Vanity Fair, Urban Sodium, and Steel Green, respectively.
    The frame-camera measurements are compared with reference spectra
    recorded using the Avantes spectrometer. Both measurements were
    corrected for the corresponding illumination spectrum. For visualization, the mean spectrum within each ROI was normalized to the range 0–1, while the corresponding spatial standard deviation was scaled by the same normalization factor and plotted as a \(\pm 1\) SD envelope around the normalized mean. Consequently, the shaded envelope may extend beyond the 0–1 range and should be interpreted solely as an indication of pixel-to-pixel spectral variability rather than as physically bounded transmission values.}
    \label{fig:camera_results}
\end{figure*}

For transmission reconstruction, the spectrum measured at each pixel was normalized to a reference spectrum acquired from a sample-free region. This procedure compensates for the wavelength-dependent spectral response of the illumination source and detection system. The reference-region normalization used here is sufficient for the present comparison of normalized spectral profiles, but it should not be interpreted as a complete radiometric calibration of the hyperspectral image. Quantitative transmittance or reflectance imaging would require a separate reference acquisition under unchanged optical conditions and normalization of each sample-pixel spectrum by the corresponding reference-pixel spectrum, thereby compensating spatial as well as spectral nonuniformities of the illumination and imaging system, as commonly performed in common-path Fourier-transform spectral imaging \cite{perri2021visible}. For comparison with
the independently measured spectrometer references, the resulting
spectra were additionally normalized to their respective maximum
values. Figure~\ref{fig:camera_results} shows the reconstructed
false-color image and the spectra extracted from the four filter
regions. The false-color image was generated by averaging the
reconstructed datacube over the blue (440-470~nm), green
(510-560~nm), and red (620-700~nm) spectral intervals.

\subsection{Derivative-Based Event Reconstruction}

In the first acquisition approach, the sample was illuminated with a LED, operating in continuous-wave mode, and the spectral scan was produced solely by continuously translating the GEMINI wedge assembly. 
For each acquisition, the wedge position was scanned from -0.4\,mm to 0.4\,mm at an operating speed of \SI{2}{\milli\meter\per\second}, while the interferometric image was recorded with the event-based camera. As the GEMINI position changes, the interference fringes move across the sensor plane, converting the spectral information of the sample into a time-varying irradiance signal at each pixel.

Unlike conventional frame-based sensors, an event camera does not directly sample absolute intensity. Instead, each pixel independently monitors changes in logarithmic intensity and generates an event
\begin{equation}
e_k=(x_k,y_k,t_k,p_k),
\end{equation}
when the change relative to its internal reference exceeds the contrast threshold. Here, $(x_k,y_k)$ denotes the pixel position, $t_k$ is the microsecond-resolution timestamp, and $p_k\in{-1,+1}$ indicates a decrease or increase in brightness, respectively. Consequently, regions with temporally varying intensities generate dense event streams, whereas regions with constant intensities produce few or no events, reducing redundant measurements.

During the continuous scanning of the GEMINI interferometer, the event stream therefore encodes changes in the interferogram rather than its absolute intensity. For a constant scan velocity $v=dx/dt$, and assuming small relative intensity variations,

\begin{equation}
E(x)\propto\frac{d}{dt}\log I(x(t))
=v\frac{d}{dx}\log I(x)
\approx \frac{dI(x)}{dx}.
\label{eq:event_derivative}
\end{equation}
The event-domain signal can thus be interpreted as a sampled derivative of the conventional interferogram.

Using the derivative property of the Fourier transform,
\begin{equation}
\mathcal{F}_x{E(x)}
\propto
i2\pi\nu,\mathcal{F}_x{I(x)},
\end{equation}
and an intensity-like spectrum can therefore be estimated as
\begin{equation}
\hat{S}(\nu)
\propto
\left|
\frac{\mathcal{F}_x{E(x)}}
{i2\pi\nu}
\right|.
\label{eq:event_spectrum}
\end{equation}

For reconstruction, the event timestamps are mapped to the synchronized GEMINI position and accumulated into position bins, forming a stack $E(x_k,y,z)$, where $x_k$ denotes the interferometer position and $(y,z)$ the sensor coordinates. For each pixel or region of interest, the resulting event-domain interferogram is background-corrected, apodized, Fourier transformed along $x_k$, and corrected according to Eq.~\ref{eq:event_spectrum} to obtain the reconstructed spectrum.

\begin{figure}[h!]
    \centering
    \includegraphics[width=0.8\linewidth]{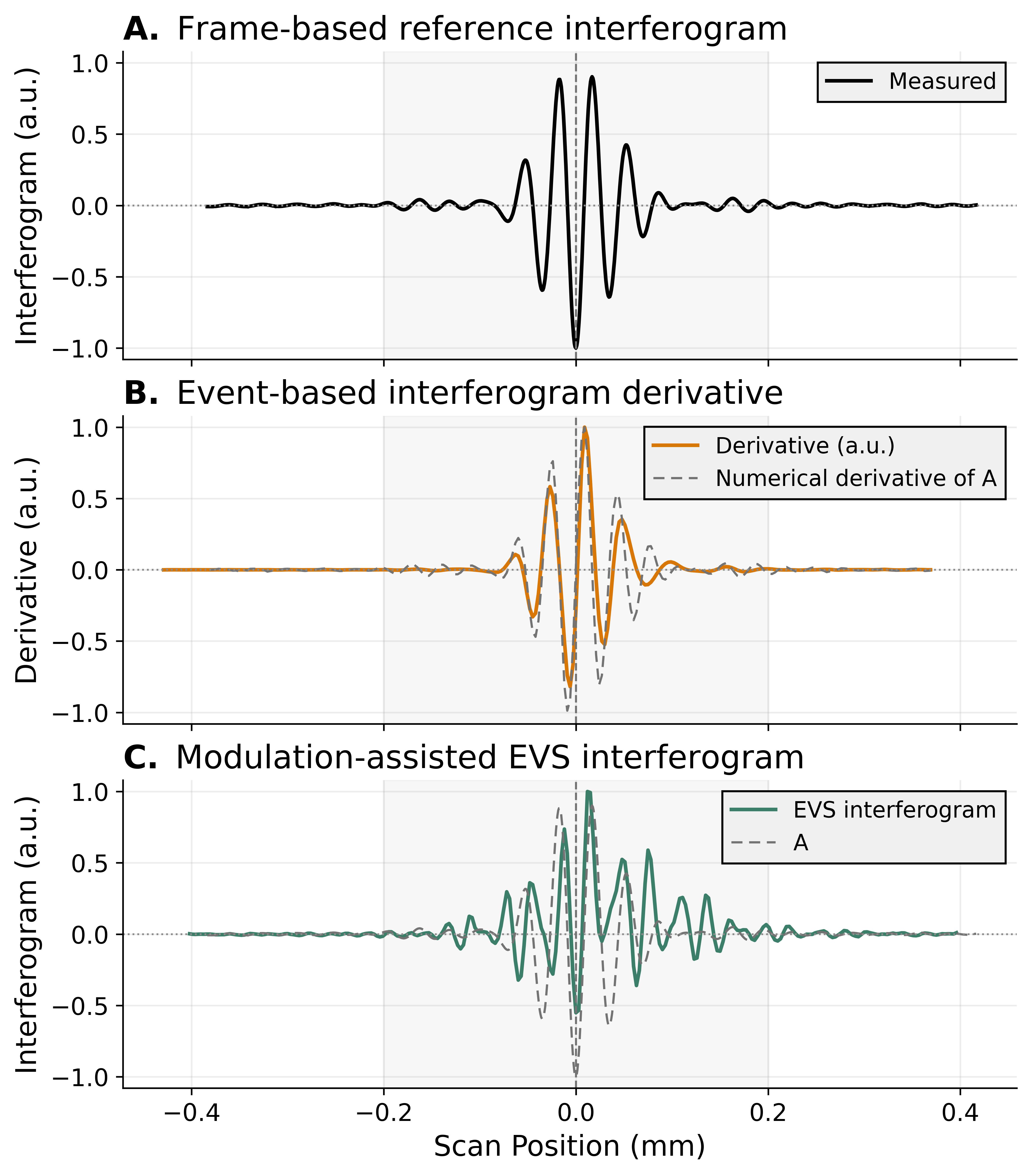}
    \caption{Comparison of interferogram acquisition modes.
    (A) Reference interferogram recorded with a conventional frame-based camera, providing the direct intensity variation as a function of GEMINI scan position. (B) Event-based acquisition under continuous GEMINI scanning, where the EVS records a derivative-like version of the interferogram. The dashed curve shows the numerical derivative of the frame-based interferogram for comparison. (C) Modulation-assisted EVS acquisition, in which the event stream more directly recovers the interferometric signal. The shaded region highlights the central high-contrast portion of the interferogram, where most of the spectral information is concentrated. Minor broadening or local shifts in the event-based traces may arise from continuous scanning of the GEMINI wedges, which are primarily designed for step-scan operation.}
    \label{fig:Interferograms}
\end{figure}

Figure~\ref{fig:Interferograms} compares results obtained with each of the two measurement regimes. Each pixel of  a conventional camera records the intensity interferogram directly, as shown in Fig.~\ref{fig:Interferograms}A, whereas the event-based camera responds primarily to changes in the interferogram, producing the derivative-like signal shown in Fig.~\ref{fig:Interferograms}B. Because the fringe contrast decreases away from the central region of the interferogram, the intensity variations can fall below the event-generation threshold. As a result, fewer events are produced in the outer regions of the scan, and the event-domain interferogram appears to flatten more rapidly than the corresponding frame-camera interferogram. 
It should also be noted that small deviations can arise from the operating mode of the GEMINI interferometer. The device is designed primarily for step-scan operation, where each wedge position includes a settling time to ensure mechanical and phase stability. In the present event-based acquisition, however, the wedge was driven continuously to generate a moving fringe pattern. Although the recorded position trace was verified to be approximately linear over the scan, continuous operation can introduce small motor-induced drifts or transient positioning errors. These effects may locally broaden or shift parts of the event-measured interferogram, explaining regions where the event-domain response appears slightly wider than expected.

\begin{figure*}[h!]
    \centering
    \includegraphics[width=\linewidth]{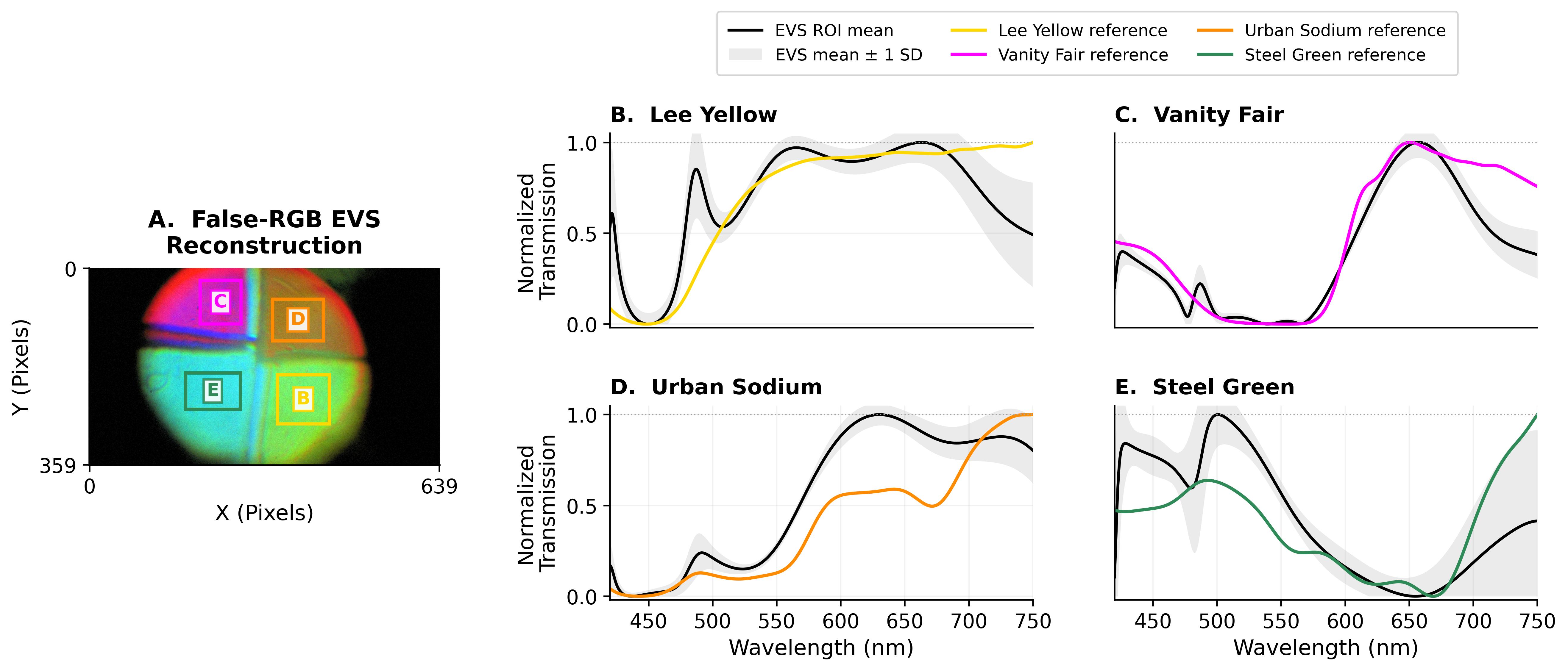}
    \caption{Derivative-based hyperspectral reconstruction of the sample.
    (A) False-color representation of the sample generated by averaging the reconstructed datacube over the blue (440-470~nm), green (510-560~nm), and red (620-700~nm) spectral intervals. The numbered markers indicate the regions from which the spectra were
    extracted. (B-E) Reconstructed transmission spectra for LEE
    Yellow, Vanity Fair, Urban Sodium, and Steel Green, respectively.
    The event-based measurements are compared with reference spectra
    recorded using the Avantes spectrometer. Both measurements were
    corrected for the corresponding illumination spectrum and
    normalized to their individual maximum values for comparison of
    their spectral profiles.}
    \label{fig:derivative_results}
\end{figure*}

The results obtained for this case study are presented in Figure \ref{fig:derivative_results}. The reconstructed hyperspectral datacube is visualized as a false-RGB image of the sample, while the corresponding spectra extracted from representative regions of the four colored filters are compared with the reference measurements acquired using the Avantes spectrometer (integration of a small area). It is possible to see that the main spectral shapes of the filters are correctly recovered, including the characteristic structure around 600-650~nm in Urban Sodium, as well as the broad long-wavelength transmission bands of Vanity Fair. Nevertheless, some deviations from the reference spectra remain. These differences are especially evident for Steel Green, where the general transmission of the illumination source is substantially weaker. In this region, noise and reconstruction artifacts can generate spurious spectral features that are subsequently amplified when the reconstructed spectra are normalized by the source spectrum. The measured illumination spectrum is included in section S1 of the Supplementary Material. Note that figure \ref{fig:Interferograms} shows that the recorded derivative signal rapidly decays toward zero as the scan moves away from the central region of the interferogram. Because the spectral resolution of the reconstructed hyperspectral datacube is determined by the effective oscillation range over which the interferogram is recorded, this rapid loss of signal may limit the reconstruction of sharp spectral transitions. Considering that the interest region of the interferogram is between -0.2 and 0.2\,mm, at a wedge translation speed of \SI{2}{\milli\meter\per\second}, this excursion requires approximately 200\,ms, corresponding to an effective hyperspectral datacube acquisition rate of approximately 5\,Hz.

Note that additional discrepancies may arise from the event-generation threshold, the reduced number of events recorded away from the central high-contrast region of the interferogram, and residual uncertainties introduced during spectral normalization and calibration. Despite these limitations, the positions and overall shapes of the dominant spectral features are preserved, demonstrating that the proposed reconstruction can distinguish the four filters through their characteristic transmission profiles. 

\subsection{Light-Modulation-Based Event Reconstruction}

To extend the useful interferometric range and improve the achievable spectral resolution, a second acquisition strategy was developed based on optical intensity modulation. In this approach, the sample was illuminated with a white LED driven by a square-wave signal at 6\,kHz. The periodic switching of the illumination generates a well-defined intensity transition at every rising and falling edge of the modulation waveform. Consequently, in this configuration, the event camera records the interference pattern reaching the sensor at each LED transition, rather than responding only to the comparatively small changes between consecutive interferometric patterns produced by the GEMINI displacement.

Once more, the GEMINI wedges were translated continuously, with displacement range of -0.4 to 0.4\,mm, resulting in an effective acquisition rate of 2.5\,fps. Thus, optical modulation preserves the acquisition-speed advantage of continuous scanning, maintaining the scanning speed but increasing the scanning range. Higher translation velocities were not investigated in this work as the interferometer motor is optimized for step-scan operation and operation at elevated velocities may introduce additional positioning uncertainties and mechanical instabilities.

\begin{figure*}[h!]
    \centering
    \includegraphics[width=0.9\linewidth]{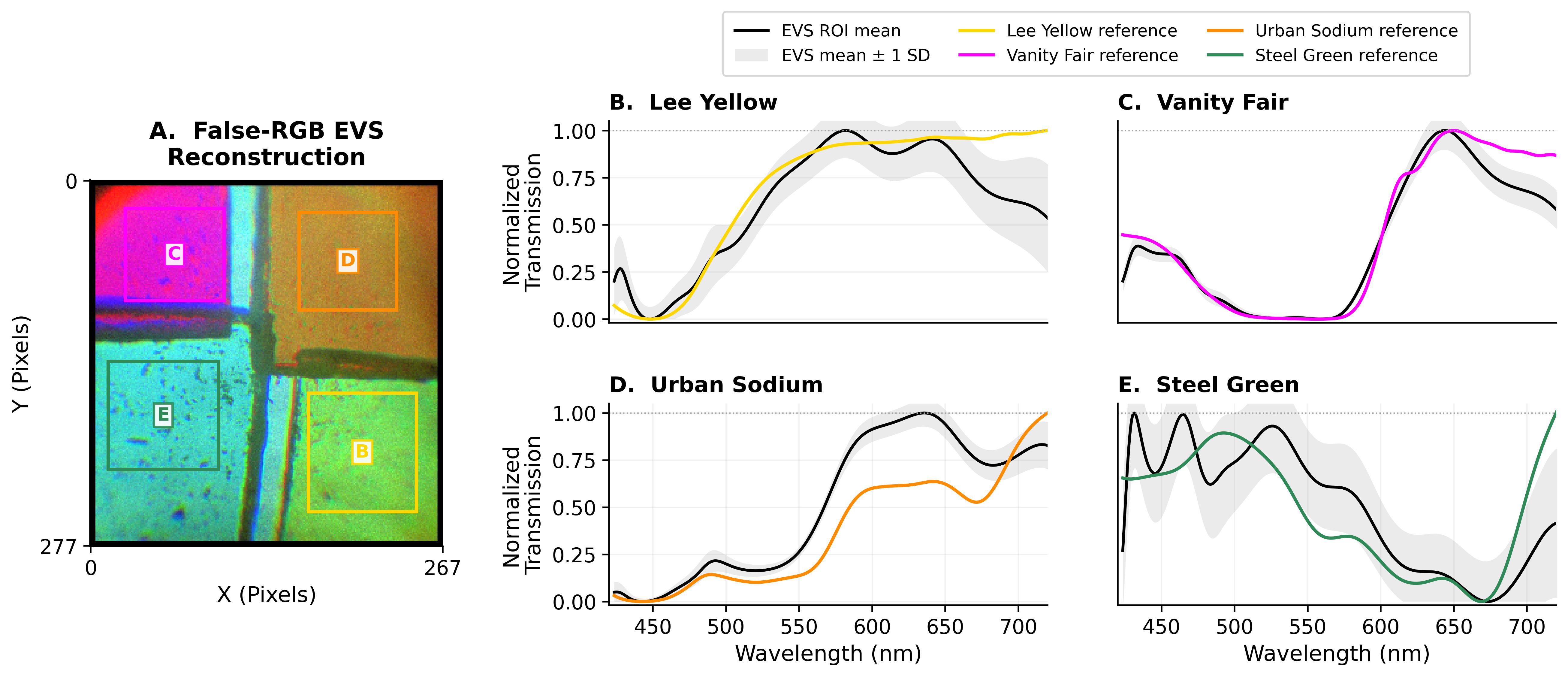}
    \caption{Light modulation-based hyperspectral reconstruction of the sample.
    (A) False-color representation of the sample generated by averaging the reconstructed datacube over the blue (440-470~nm), green (510-560~nm), and red (620-700~nm) spectral intervals. The numbered markers indicate the regions from which the spectra were
    extracted. (B-E) Reconstructed transmission spectra for selected regions corresponding to the LEE
    Yellow, Vanity Fair, Urban Sodium, and Steel Green, respectively. The black curve shows the regional mean spectrum, and the gray shaded envelope represents the pixel-wise minimum-to-maximum spectral range.
    The event-based measurements are compared with reference spectra
    recorded using the Avantes spectrometer. Both measurements were
    corrected for the corresponding illumination spectrum and
    normalized to their individual maximum values for comparison of
    their spectral profiles.}
    \label{fig:led_results}
\end{figure*}

When the LED is switched on, each pixel compares the illuminated interferometric intensity with the low-intensity state recorded while the source is switched off. Conversely, the falling edge produces events of the opposite polarity as the signal returns to the dark state, both transitions therefore provide samples of the instantaneous interferometric pattern. In this configuration, the event response is referenced to an approximately constant dark baseline rather than to the preceding GEMINI position. The accumulated event signal can consequently be treated as an intensity-like interferogram, rather than as its derivative, as can be seen in Figure \ref{fig:Interferograms}C. The derivative compensation used in the previous reconstruction method is therefore no longer required, and the spectrum can be recovered directly by Fourier transformation of the interferogram. Similar to the previous case, the reconstructed image and filter spectra for this approach are represented in Figure \ref{fig:led_results}, using the same layout of the reference case.

Compared with the derivative-based reconstruction in Fig.~\ref{fig:derivative_results}, the modulation-assisted approach generally provides closer agreement with the reference transmission spectra while preserving the same main spectral features. For Lee Yellow, the modulation-based reconstruction more closely follows the rising transmission edge and broad high-transmission region, although the reconstructed signal still decreases more strongly than the reference toward longer wavelengths. Vanity Fair exhibits the closest agreement between the two event-based approaches and the spectrometer reference, with the low-transmission region and the sharp transition near $600~\mathrm{nm}$ accurately reproduced. As in the derivative-based case, the reconstructed transmission decreases somewhat more rapidly at the longest wavelengths. For Urban Sodium, both approaches recover the overall increase toward the red and the local decrease near $680~\mathrm{nm}$, while the modulation-based reconstruction provides a closer representation of the reference spectral shape in the $580$--$660~\mathrm{nm}$ interval, despite an overestimation of the transmission. Steel Green remains the most challenging case for the event-based reconstructions. Both methods recover the broad decrease toward the minimum near $670$--$680~\mathrm{nm}$, but only the modulation approach recovers the oscillations around $580$ and $650$\,nm. Still, the reconstructed spectra exhibit substantial deviations at shorter wavelengths. Since shorter wavelengths correspond to higher interferometric spatial frequencies, this spectral region is inherently more sensitive to finite sampling, positioning uncertainties, and nonuniformities in the scan, and would be the first to exhibit aliasing if the effective sampling approached the Nyquist limit. The gray envelopes, representing the regional mean $\pm 1$ standard deviation, show that the reconstructed spectra are spatially consistent over much of the measured range, with increased variability predominantly occurring at higher wavelengths. Overall, the modulation-assisted acquisition improves the spectral agreement relative to the derivative-based approach while directly recovering an intensity-like interferogram without requiring derivative compensation. In addition, the fixed modulation frequency provides a well-defined temporal sampling interval. Under the approximately constant GEMINI translation velocity, this corresponds to a nearly uniform sampling step along the interferometric coordinate, improving the consistency of the interferogram sampling.


Generally, the results obtained with the modulation-assisted approach, particularly in the $550$--$650~\mathrm{nm}$ range, suggest an improved spectral resolution compared with the derivative-based method. To experimentally assess this difference, both acquisition schemes were used to reconstruct the spectrum of a narrowband $532~\mathrm{nm}$ laser, as detailed in Section S2 of the Supplementary Material. The modulation-assisted reconstruction yielded a FWHM of $17.9~\mathrm{nm}$ when using the full $-0.4$ to $+0.4~\mathrm{mm}$ interferometer range (modulated version), compared with $39.2~\mathrm{nm}$ for the derivative-based reconstruction. When the modulation-assisted dataset was restricted to the same $-0.2$ to $+0.2~\mathrm{mm}$ range, its FWHM increased to $35.4~\mathrm{nm}$, confirming that the improved resolution observed for the full-range modulation-assisted measurement is primarily enabled by the larger sampled interferometer range. While the intrinsic linewidth of the laser was not independently known, the reference spectrometer measured a substantially narrower FWHM of $2.06~\mathrm{nm}$, indicating that the broader linewidths obtained with the event-based reconstructions are predominantly determined by the effective spectral resolution of the interferometric measurement rather than by the laser emission itself.

\section{Discussion}

The two event-based acquisition strategies rely on the same continuous translation of the GEMINI wedges, but they differ fundamentally in the intensity reference used to generate events. In the derivative-based approach, each pixel compares the current interferometric intensity with the level stored after the preceding event. The resulting signal therefore follows the local temporal variation of the interferogram and can be interpreted as a derivative-like measurement along the scan coordinate. Recovering an intensity-equivalent spectrum consequently requires compensation in the Fourier domain through division by the interferometric spatial frequency. Although this correction enables the principal spectral signatures to be reconstructed, it also increases sensitivity to noise and to departures from the ideal derivative model. In addition, as the fringe contrast decreases away from the central interferogram, the corresponding intensity variations may fall below the event-generation threshold. The useful interferometric range is therefore reduced, limiting the amount of information available for resolving narrower spectral features.

The modulation-based approach changes this measurement principle by periodically switching the LED between an illuminated state and an approximately dark reference state. At each rising or falling transition, the intensity difference observed by a pixel is dominated by the instantaneous illuminated interferometric pattern rather than by the comparatively small change produced between neighboring GEMINI positions. This stronger and more consistent reference improves the recovery of low-contrast interferometric oscillations and avoids the derivative compensation required by the first approach. The improved agreement observed in Fig.~\ref{fig:led_results} is therefore consistent with the modulation scheme preserving the interferometric intensity information more directly. Nevertheless, the performance of this implementation remains influenced by the spectral output of the selected LED, particularly near the limits of its emission range. This restriction is associated with the illumination source used in the experiment rather than with the modulation principle itself, which could be extended to other illumination sources.


To quantify the agreement between the reconstructed and reference
spectra, the Pearson correlation was calculated for each filter. The
comparison was performed over the common wavelength interval sampled
by both measurements. For each filter, the EVS spectrum corresponded
to the spatially averaged spectrum of the selected region, while the
reference spectrum was obtained using the spectrometer.

Before calculating the metrics, both spectra were independently normalized
to their maximum value:

\begin{equation}
\widetilde{S}_{j}(\lambda_i)
=
\frac{S_{j}(\lambda_i)}
{\max_{\lambda} S_{j}(\lambda)},
\qquad
j \in \{\mathrm{EVS},\mathrm{ref}\},
\end{equation}

where $\lambda_i$ denotes the $i$th wavelength sample and $N$ is the
total number of wavelength samples included in the comparison.

\begin{figure}[h!]
    \centering
    \includegraphics[width=0.9\linewidth]{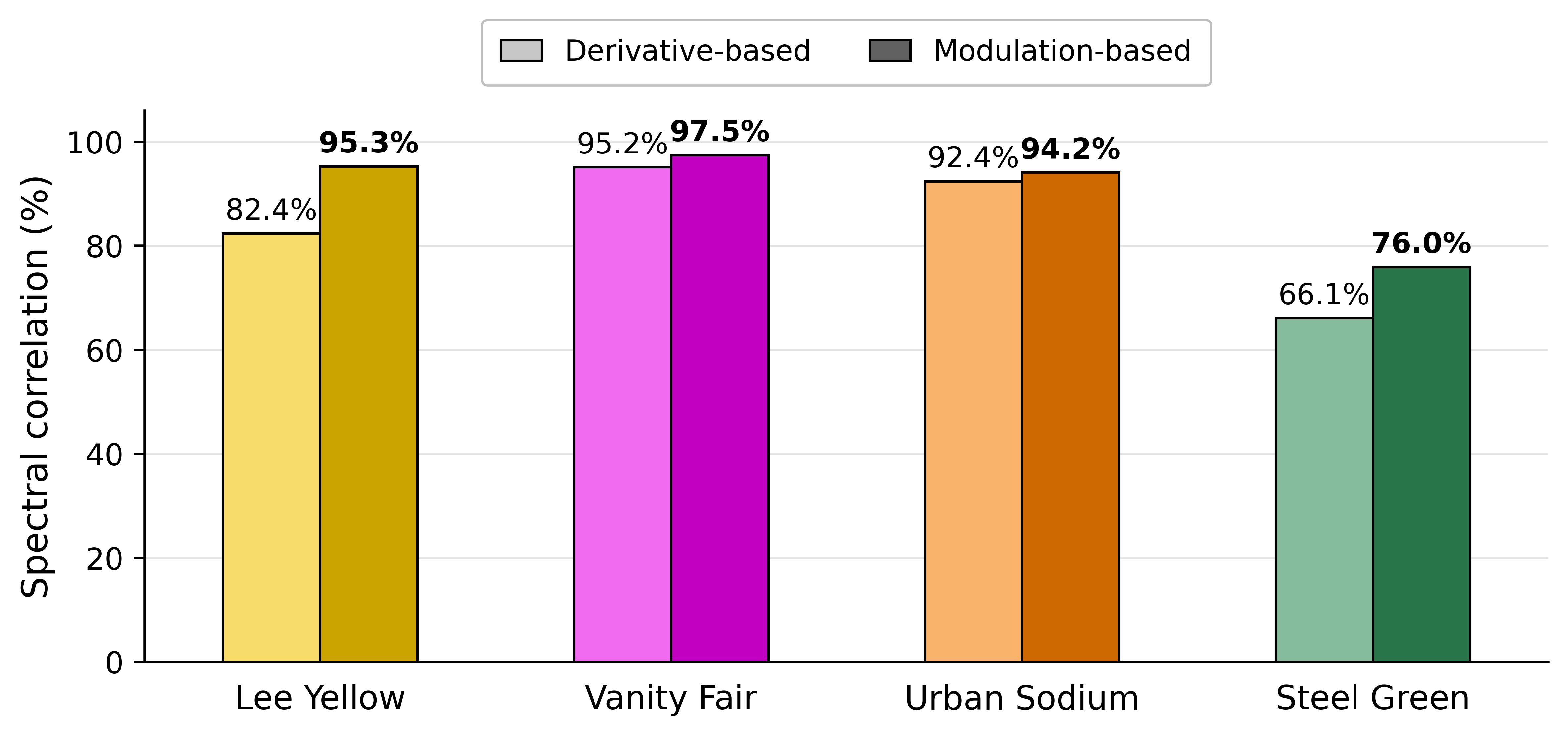}
    \caption{Quantitative comparison of the derivative-based and modulation-based event reconstructions. Pearson correlation coefficient between the reconstructed transmission spectra and the corresponding spectrometer references for each filter. Values are expressed as percentages. Higher correlation indicates better agreement.}
    \label{fig:metrics}
\end{figure}

The Pearson correlation coefficient was calculated as

\begin{equation}
r =
\frac{
\sum_{i=1}^{N}
\left(
\widetilde{S}_{\mathrm{EVS},i}
-
\overline{\widetilde{S}}_{\mathrm{EVS}}
\right)
\left(
\widetilde{S}_{\mathrm{ref},i}
-
\overline{\widetilde{S}}_{\mathrm{ref}}
\right)
}{
\sqrt{
\sum_{i=1}^{N}
\left(
\widetilde{S}_{\mathrm{EVS},i}
-
\overline{\widetilde{S}}_{\mathrm{EVS}}
\right)^2
}
\sqrt{
\sum_{i=1}^{N}
\left(
\widetilde{S}_{\mathrm{ref},i}
-
\overline{\widetilde{S}}_{\mathrm{ref}}
\right)^2
}
},
\end{equation}

where the overbars indicate the mean value over the evaluated
wavelength range. The correlation coefficient measures the similarity
between the spectral profiles, with $r=1$ indicating identical
spectral shapes up to a linear scaling. For presentation
in Fig.~\ref{fig:metrics}, the $r$ values were multiplied by 100 and expressed as percentages. 

The correlation coefficients provide a quantitative assessment of the agreement between the reconstructed and reference spectral shapes. The mean correlation increased from approximately 84\% for the derivative-based approach to 91\% for the modulation-assisted method, with higher correlation values obtained for all four filters. These results indicate that both approaches preserve the principal spectral signatures of the filters, while the modulation-assisted acquisition provides consistently closer agreement with the reference spectra.

Beyond acquisition speed, the event-based representation remains intrinsically sparse because only pixels undergoing sufficient interferometric intensity variation generate events. After binning by interferometer position, both representations consisted predominantly of zero-valued entries, accounting for 94.2\% of the derivative-based representation and 82.7\% of the modulation-assisted representation. A detailed comparison of the numerical representation, bit depth, and sparsity of the frame- and event-based datasets is provided in Section S3 of the Supplementary Material. The modulation-assisted strategy therefore trades increased event density for improved spectral fidelity.

To further demonstrate the applicability of the proposed framework beyond controlled transmission measurements, the derivative-based acquisition was also applied to a microplastic particle imaged in reflection under continuous illumination. A complete hyperspectral acquisition was obtained in approximately $200~\mathrm{ms}$, and the reconstructed blue, green, and red spectral bands were combined into a false-RGB representation that preserved the principal morphology and color variation of the particle. Moreover, the event stream was strongly concentrated around the sample, with $90\%$ of all recorded events originating from only $24.23\%$ of the sensor area, illustrating the spatial sparsity naturally provided by event-based sensing for localized reflective objects. Further details and the corresponding reconstructed spectral images are provided in Section S4 of the Supplementary Material.

We further explored the event stream's ability to retain spatial information during continuous acquisition using a moving reflective particle. As detailed in Section S5 of the Supplementary Material, event-count images were used to estimate the particle trajectory during the interferometric scan, after which the signed-event stack was transformed into a co-moving reference frame before spectral reconstruction. The resulting hyperspectral image preserved the spatial morphology of the particle and reproduced its independently measured spectral profile with a Pearson correlation of approximately 93\%. Although the present experiment employed a comparatively slow scan because of the limited reflected optical power, it demonstrates that spatial motion can be compensated before Fourier reconstruction, extending the proposed framework toward hyperspectral imaging of dynamic scenes.

Note that the present implementation remains limited primarily by the interferometer and by the operating conditions used to generate measurable event contrast. Motor translation velocities above \SI{2}{\milli\meter\per\second} were not investigated to avoid imposing additional stress on the GEMINI motor, which was designed principally for step-scan operation. The demonstrated 5\,Hz acquisition rate should therefore be regarded as a system-level proof of concept rather than as the temporal limit of the event sensor. A translation mechanism designed for continuous high-speed scanning could support shorter acquisition times and improve position repeatability.

A limitation of the present event-based implementation is that the reconstructed spectral amplitudes should not be interpreted as absolute optical intensity, transmittance, or reflectance. Event cameras do not directly measure intensity but generate events when changes in logarithmic intensity exceed a contrast threshold. Consequently, the accumulated event signal depends not only on the underlying optical intensity variation but also on the pixel-wise contrast thresholds, event quantization, and the previous internal reference state. The derivative-based reconstruction further approximates this event signal as the derivative of the interferogram and compensates it in the Fourier domain. Although this procedure reliably recovers the wavelength positions and relative shapes of the dominant spectral features, it does not preserve a calibrated linear relationship between reconstructed spectral amplitude and incident optical power. Similar limitations arise in other event-based hyperspectral approaches, for which recovery of absolute intensity requires additional reference information and sensor calibration \cite{chen2026self}. Because the objective here is the recovery of relative spectral signatures rather than absolute transmittance, a sample-free region recorded within the same event acquisition can be used as the spectral reference. This simultaneous reference experiences the same illumination, interferometer scan, detector settings, and acquisition timing as the sample regions, allowing common wavelength-dependent system responses to be suppressed without implying an absolute radiometric calibration.
Accordingly, the present quantitative analysis focuses on spectral shape rather than absolute radiometric amplitude. Reconstructed and reference spectra are independently normalized before comparison, and Pearson correlation is used to quantify agreement between their wavelength-dependent profiles.

\section{Conclusion}
We have demonstrated a high-speed event-based Fourier-transform hyperspectral microscopy framework that combines an HD event camera with a common-path birefringent interferometer. By continuously translating the interferometer wedges, the spectral information encoded in the evolving interference pattern is transferred into an asynchronous event stream without requiring spatial multiplexing, spectrally swept illumination, or repeated full-frame readout. Two complementary reconstruction strategies were investigated. Under continuous illumination, the event signal behaves as a derivative-like interferogram and can be reconstructed using Fourier-domain derivative compensation. With 6\,kHz intensity modulation, the events instead provide a more direct intensity-like interferometric measurement, extending the useful oscillation range and improving the recovery of sharp spectral structures.

Both approaches recovered the principal transmission features of four photographic filters and produced spatially resolved false-RGB reconstructions. Under the present operating conditions, the useful event-domain scan interval was acquired in approximately 200\,ms, corresponding to an effective acquisition rate of approximately 5 hyperspectral datacubes per second, compared with approximately 15\,min for the reference frame-based step scan. The event-domain representations also required lower accumulated-value bit depth and contained substantially fewer nonzero bins than the dense frame stack, although the exact end-to-end storage benefit depends on the encoding overhead of the individual events.

Finally, reflection-mode imaging of a microplastic particle provided a proof of concept beyond the controlled filter measurements. The experiment recovered spatially varying blue, green, and red responses within a 200\,ms acquisition, while 90\% of the recorded events were concentrated within only 24.23\% of the sensor area. This result is not presented as polymer identification, but as evidence that the proposed framework can recover spatially resolved spectral contrast from a heterogeneous sample while naturally suppressing redundant measurements from an inactive background. Overall, the results establish event-based detection as a promising route for transforming Fourier-transform hyperspectral microscopy from dense frame-by-frame sampling into a continuous, asynchronous, and spatially selective acquisition process. Future developments based on faster continuous-scan interferometers, broadband modulated illumination, improved optical throughput, radiometric calibration, and real-time motion compensation could extend the demonstrated dynamic proof of concept toward quantitative high-speed hyperspectral microscopy of complex moving scenes.

\begin{backmatter}
\bmsection{Funding} 
This work was co-financed by the National Funds through the FCT—Fundação para a Ciência e a Tecnologia, I.P. (Portuguese Foundation for Science and Technology) within the project HyperQuIm, with reference 2024.17596.PEX (https://doi.org/10.54499/2024.17596.PEX). This work was co-financed by National Funds through the FCT - Fundação para a Ciência e a Tecnologia, I.P. (Portuguese Foundation for Science and Technology) within the project QuAC, with reference 2024.16086.PEX (https://doi.org/10.54499/2024.16086.PEX).

\bmsection{Acknowledgment} 
Joana Teixeira acknowledges the support of FCT - Fundação para a Ciência e Tecnologia, I.P. by project reference 2024.00426.BD and DOI identifier https://doi.org/10.54499/2024.00426.BD.  Tomás Lopes acknowledges the support of FCT - Fundação para a Ciência e Tecnologia, I.P. by project reference 2024.01830.BD, and DOI identifier https://doi.org/10.54499/2024.01830.BD. Tiago D. Ferreira acknowledges the support of FCT under the grant 2024.10684.CEECIND.

\bmsection{Disclosures} The authors declare no conflicts of interest.

\bmsection{Data availability} Data underlying the results presented in this paper are not publicly available at this time but may be obtained from the authors upon reasonable request.

\bigskip

\bmsection{Supplemental document}
See Supplementary Material for the supporting content mentioned along the manuscript.

\end{backmatter}

\bibliography{sample}

@inproceedings{prabhavathy2019unsupervised,
  title={Unsupervised learning method for mineral identification from hyperspectral data},
  author={Prabhavathy, P and Tripathy, BK and Venkatesan, M},
  booktitle={International Conference on Innovations in Bio-Inspired Computing and Applications},
  pages={148--160},
  year={2019},
  organization={Springer}
}

@article{maktabi2025detection,
  title={Detection of flap malperfusion after microsurgical tissue reconstruction using hyperspectral imaging and machine learning},
  author={Maktabi, Marianne and Huber, Benjamin and Pfeiffer, Toni and Schulz, Torsten},
  journal={Scientific Reports},
  volume={15},
  number={1},
  pages={15637},
  year={2025},
  publisher={Nature Publishing Group UK London}
}

@article{giannoni2026hyperspectral,
  title={Hyperspectral imaging solutions for brain tissue metabolic and haemodynamic monitoring: an updated perspective},
  author={Giannoni, Luca and Lange, Fr{\'e}d{\'e}ric and Tachtsidis, Ilias},
  journal={Journal of Optics},
  volume={28},
  number={6},
  pages={062501},
  year={2026},
  publisher={IOP Publishing}
}

@article{pahlawan2025non,
  title={Non-destructive identification of microplastics in soil using spectroscopy and hyperspectral imaging},
  author={Pahlawan, MFR and Kim, Yena and Aline, Umuhoza and Zahroh, Aridatuz and Masithoh, Rudiati Evi and Kim, Moon S and Baek, Insuck and Cho, Byoung-Kwan},
  journal={TrAC Trends in Analytical Chemistry},
  volume={187},
  pages={118216},
  year={2025},
  publisher={Elsevier}
}

@article{zhang2025monitoring,
  title={Monitoring soil salinization in Arid cotton fields using Unmanned Aerial Vehicle hyperspectral imagery},
  author={Zhang, Jinming and Ding, Jianli and Tan, Jiao and Wang, Jinjie and Zhang, Zihan and Wang, Zeyuan and Ge, Xiangyu},
  journal={International Journal of Applied Earth Observation and Geoinformation},
  volume={140},
  pages={104584},
  year={2025},
  publisher={Elsevier}
}

@article{hong2026hyperspectral,
  title={Hyperspectral imaging},
  author={Hong, Danfeng and Li, Chenyu and Yokoya, Naoto and Zhang, Bing and Jia, Xiuping and Plaza, Antonio and Gamba, Paolo and Benediktsson, Jon Atli and Chanussot, Jocelyn},
  journal={Nature Reviews Methods Primers},
  volume={6},
  number={1},
  pages={19},
  year={2026},
  publisher={Nature Publishing Group UK London}
}

@article{Sinclair_06,
author = {Michael B. Sinclair and David M. Haaland and Jerilyn A. Timlin and Howland D. T. Jones},
journal = {Appl. Opt.},
number = {24},
pages = {6283--6291},
publisher = {Optica Publishing Group},
title = {Hyperspectral confocal microscope},
volume = {45},
month = {Aug},
year = {2006},
url = {https://opg.optica.org/ao/abstract.cfm?URI=ao-45-24-6283},
doi = {10.1364/AO.45.006283},

}

@article{chabrillat2024enmap,
  title={The EnMAP spaceborne imaging spectroscopy mission: Initial scientific results two years after launch},
  author={Chabrillat, Sabine and Foerster, Saskia and Segl, Karl and Beamish, Alison and Brell, Maximilian and Asadzadeh, Saeid and Milewski, Robert and Ward, Kathrin J and Brosinsky, Arlena and Koch, Katrin and others},
  journal={Remote Sensing of Environment},
  volume={315},
  pages={114379},
  year={2024},
  publisher={Elsevier}
}

@article{holman2020autonomous,
  title={Autonomous adaptive data acquisition for scanning hyperspectral imaging},
  author={Holman, Elizabeth A and Fang, Yuan-Sheng and Chen, Liang and DeWeese, Michael and Holman, Hoi-Ying N and Sternberg, Paul W},
  journal={Communications biology},
  volume={3},
  number={1},
  pages={684},
  year={2020},
  publisher={Nature Publishing Group UK London}
}

@article{von2025image,
  title={Image stitching for real-time laparoscopic hyperspectral imaging},
  author={von Braun, Marie-Sophie and Pfahl, Annekatrin and Melzer, Andreas and Chalopin, Claire and K{\"o}hler, Hannes},
  journal={Scientific Reports},
  volume={15},
  number={1},
  pages={28010},
  year={2025},
  publisher={Nature Publishing Group UK London}
}

@article{gao2011snapshot,
  title={Snapshot hyperspectral retinal camera with the Image Mapping Spectrometer (IMS)},
  author={Gao, Liang and Smith, R Theodore and Tkaczyk, Tomasz S},
  journal={Biomedical optics express},
  volume={3},
  number={1},
  pages={48--54},
  year={2011},
  publisher={Optical Society of America}
}

@inproceedings{puckrin2009airborne,
  title={Airborne measurements in the infrared using FTIR-based imaging hyperspectral sensors},
  author={Puckrin, E and Turcotte, CS and Lahaie, P and Dub{\'e}, D and Lagueux, P and Farley, V and Marcotte, F and Chamberland, M},
  booktitle={Electro-Optical Remote Sensing, Photonic Technologies, and Applications III},
  volume={7482},
  pages={196--207},
  year={2009},
  organization={SPIE}
}

@article{lichtsteiner2008,
  title={A 128x128 120 dB 15$\mu$s latency asynchronous temporal contrast vision sensor},
  author={Lichtsteiner, Patrick and Posch, Christoph and Delbruck, Tobi},
  journal={IEEE journal of solid-state circuits},
  volume={43},
  number={2},
  pages={566--576},
  year={2008},
  publisher={IEEE}
}

@article{gallego2020,
  title={Event-based vision: A survey},
  author={Gallego, Guillermo and Delbr{\"u}ck, Tobi and Orchard, Garrick and Bartolozzi, Chiara and Taba, Brian and Censi, Andrea and Leutenegger, Stefan and Davison, Andrew J and Conradt, J{\"o}rg and Daniilidis, Kostas and others},
  journal={IEEE transactions on pattern analysis and machine intelligence},
  volume={44},
  number={1},
  pages={154--180},
  year={2020},
  publisher={IEEE}
}

@inproceedings{yu2025active,
  title={Active hyperspectral imaging using an event camera},
  author={Yu, Bohan and Liang, Jinxiu and Wang, Zhuofeng and Fan, Bin and Subpa-asa, Art and Shi, Boxin and Sato, Imari},
  booktitle={Proceedings of the Computer Vision and Pattern Recognition Conference},
  pages={929--939},
  year={2025}
}

@article{chen2026self,
  title={Self-calibrated neuromorphic hyperspectral derivative imaging},
  author={Chen, Rongzhou and Wang, Chutian and Li, Yuxing and Cao, Yuqing and Zhu, Shuo and Lam, Edmund Y},
  journal={Optica},
  volume={13},
  number={4},
  pages={587--590},
  year={2026},
  publisher={Optica Publishing Group}
}

@article{teixeira2026asynchronous,
  title={Asynchronous event-based spectroscopy for microsecond-resolved spectral reconstruction},
  author={Teixeira, Joana M and Lopes, Tomas and Ferreira, Tiago D and Monteiro, Catarina S and Jorge, Pedro AS and Silva, Nuno A},
  journal={Sensors and Actuators A: Physical},
  pages={118184},
  year={2026},
  publisher={Elsevier}
}

@article{guo2024eventlfm,
  title={EventLFM: event camera integrated Fourier light field microscopy for ultrafast 3D imaging},
  author={Guo, Ruipeng and Yang, Qianwan and Chang, Andrew S and Hu, Guorong and Greene, Joseph and Gabel, Christopher V and You, Sixian and Tian, Lei},
  journal={Light: Science \& Applications},
  volume={13},
  number={1},
  pages={144},
  year={2024},
  publisher={Nature Publishing Group UK London}
}

@article{zhang2024event,
  title={Event-based X-ray imager with ghosting-free scintillator film},
  author={Zhang, Ao and Pang, Jincong and Wu, Haodi and Tan, Qingwen and Zheng, Zhiping and Xu, Ling and Tang, Jiang and Niu, Guangda},
  journal={Optica},
  volume={11},
  number={5},
  pages={606--611},
  year={2024},
  publisher={Optica Publishing Group}
}

@article{geckeler2026event,
  title={Event Spectroscopy: Event-based Multispectral and Depth Sensing using Structured Light},
  author={Geckeler, Christian and Neugebauer, Niklas and Muglikar, Manasi and Scaramuzza, Davide and Mintchev, Stefano},
  journal={IEEE Robotics and Automation Letters},
  volume={11},
  number={3},
  pages={2658--2665},
  year={2026},
  publisher={IEEE}
}

@article{baird2025dynamic,
  title={Dynamic spectral fluorescence microscopy via event-based \& CMOS image-sensor fusion},
  author={Baird, Richard G and Majumder, Apratim and Menon, Rajesh},
  journal={Optics Express},
  volume={33},
  number={2},
  pages={2169--2178},
  year={2025},
  publisher={Optica Publishing Group}
}

@article{perri2021visible,
  title={Visible and near-infrared Fourier transform spectroscopy with a common-path interferometer},
  author={Perri, Antonio},
  journal={Journal of Physics B: Atomic, Molecular and Optical Physics},
  volume={54},
  number={11},
  pages={113001},
  year={2021},
  publisher={IOP Publishing}
}

@misc{leeFiltersColourEffect,
  author       = {{LEE Filters}},
  title        = {{Colour Effect Lighting Filters}},
  howpublished = {\url{https://leefilters.com/lighting/colour-effect-lighting-filters/}},
  note         = {Accessed: 2026-07-16}
}

@patent{ritter2024hyperspectral,
  author       = {Ritter, Scott},
  title        = {Systems, Devices, and Methods for Hyperspectral Imaging},
  number       = {US 11,860,034 B2},
  assignee     = {Raytheon BBN Technologies Corp.},
  year         = {2024},
  month        = jan,
  day          = {2},
  note         = {U.S. Patent 11,860,034}
}


\ifthenelse{\equal{\journalref}{aop}}{%
\section*{Author Biographies}
\begingroup
\setlength\intextsep{0pt}
\begin{minipage}[t][6.3cm][t]{1.0\textwidth} 
  \begin{wrapfigure}{L}{0.25\textwidth}
    \includegraphics[width=0.25\textwidth]{john_smith.eps}
  \end{wrapfigure}
  \noindent
  {\bfseries John Smith} received his BSc (Mathematics) in 2000 from The University of Maryland. His research interests include lasers and optics.
\end{minipage}
\begin{minipage}{1.0\textwidth}
  \begin{wrapfigure}{L}{0.25\textwidth}
    \includegraphics[width=0.25\textwidth]{alice_smith.eps}
  \end{wrapfigure}
  \noindent
  {\bfseries Alice Smith} also received her BSc (Mathematics) in 2000 from The University of Maryland. Her research interests also include lasers and optics.
\end{minipage}
\endgroup
}{}

\end{document}